\documentclass[aps,prc,superscriptaddress,showpacs,floatfix,nofootinbib,notitlepage,twocolumn]{revtex4-1}
\usepackage{amsmath,graphicx,float,hyperref,upgreek}

\begin{document}

\title{
%Understanding the centrality dependence of the pseudorapidity distribution in ultrarelativistic proton-nucleus collisions}
Multiplicity fluctuations and rapidity correlations in ultracentral proton-nucleus collisions}
\author{Rupam Samanta}
\affiliation{AGH University of Science and Technology, Faculty of Physics and
Applied Computer Science, aleja Mickiewicza 30, 30-059 Cracow, Poland}
\author{Jean-Yves Ollitrault}
\affiliation{Universit\'e Paris Saclay, CNRS, CEA, Institut de physique th\'eorique, 91191 Gif-sur-Yvette, France}

\begin{abstract} 
A collision between a proton and a heavy nucleus at ultrarelativistic energy creates particles whose rapidity distribution is asymmetric, with more particles emitted in the direction of the nucleus than in the direction of the proton. 
This asymmetry becomes more pronounced as the centrality estimator, defined from the energy deposited in a  calorimeter, increases. 
We argue that for high-multiplicity collisions, the variation of the impact parameter plays a negligible role, and that the fluctuations of the multiplicity and of the centrality estimator are dominated by quantum fluctuations, whose probability distribution can be well approximated by a correlated gamma distribution. 
We show that this simple model reproduces existing data, and we make quantitative predictions for collisions in the $0-0.1\%$ and $0-0.01\%$ centrality windows. 
We argue that by repeating the same analysis with a different centrality estimator, one can obtain direct information about the rapidity decorrelation in particle production. 
\end{abstract}

\maketitle
Collisions between protons and/or atomic nuclei at ultrarelativistic energies create a large number of particles. 
The particle multiplicity fluctuates event to event, and   a high-multiplicity event typically has more particles at all rapidities. 
This phenomenon, which is referred to as a long-range rapidity correlation, is deeply rooted in the dynamics of strong interactions at the level of elementary particles. 
Strong interactions at high energies involve, at an intermediate stage, longitudinally-extended objects called strings~\cite{Artru:1974hr,Andersson:1983ia,Christiansen:2015yqa} or flux tubes \cite{Casher:1978wy,Bialas:1984ye,Dumitru:2008wn,Dusling:2009ni,Broniowski:2015oif}. 
Since these intermediate objects cover a wide rapidity range, the production of particles from their decay is strongly correlated across rapidity.
In addition, a single collision between hadrons~\cite{Capella:1978rg} or nuclei~\cite{Olszewski:2013qwa} involves several  such elementary processes.  
Their number fluctuates event to event, which is often referred to as a ``volume'' fluctuation~\cite{Vechernin:2012qim}. 
A volume fluctuation creates more particles at all rapidities, and also generates long-range rapidity correlations~\cite{Broniowski:2015oif,Rohrmoser:2019xis}. 
Experimentally, these correlations have been analyzed mostly through forward-backward correlations~\cite{PHOBOS:2006mfc,STAR:2009goo,ALICE:2015kal,Raniwala:2016ugm,Bozek:2010vz,Bzdak:2012tp,Jia:2015jga,ATLAS:2016rbh,Rohrmoser:2018shp}.

\begin{figure}[h]
\begin{center}
\includegraphics[width=\linewidth]{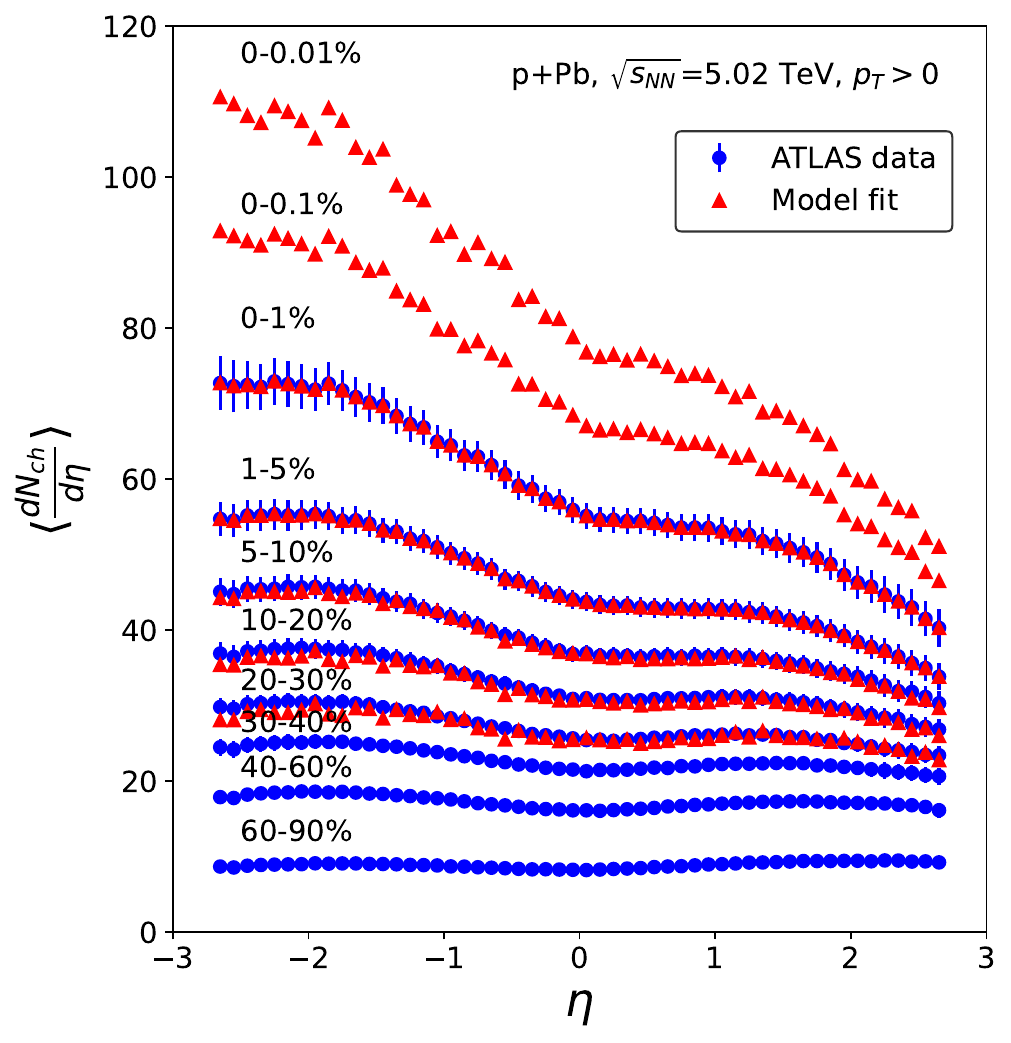} 
\end{center}
\caption{
Average multiplicity per unity pseudorapidity $\eta$ in several centrality windows in p+Pb collisions at $\sqrt{s_{\rm NN}}=5.02$~TeV per nucleon pair. 
The centrality classifier is the transverse energy $E_T$ deposited in a calorimeter covering the acceptance window $-4.9<\eta<-3.1$ in the Pb-going direction. 
Circles: ATLAS data~\cite{ATLAS:2015hkr} 
Triangles: Model calculation using Eq.~(\ref{gammaav2}), where the parameters are adjusted so as to match the data in the $0-1\%$ and $1-5\%$ centrality windows. 
}
\label{fig:etadep}
\end{figure} 

We study long-range correlations in p+Pb collisions from a different perspective, by interpreting the so-called ``centrality dependence'' of the rapidity distribution of produced particles, which has been analyzed by ATLAS~\cite{ATLAS:2015hkr} (Fig.~\ref{fig:etadep}).\footnote{Related analyses have been carried out by ALICE~\cite{ALICE:2022imr} and CMS \cite{CMS:2018xfv}.} 
%Long-range rapidity correlations in p+Pb collision have been studied previously based on hydrodynamics~\cite{Bozek:2012gr} and transport models~\cite{Ma:2014pva}. 
This analysis is done in two steps. 
First, they measure the transverse energy $E_T$ in a backward calorimeter, in the direction of the Pb nucleus, and classify events in bins of $E_T$. 
Collisions with larger $E_T$ are referred to as  ``more central'', and the bins of $E_T$ as ``centrality bins''. 
Then they measure, in each bin, the average number of charged particles as a function of the angle  $\theta$ between their direction of motion and the collision axis, or equivalently, the pseudorapidity $\eta\equiv -\ln\tan\frac{\theta}{2}$, as shown in Fig.~\ref{fig:etadep}. 
As the centrality percentile (or fraction) decreases, corresponding to a larger $E_T$, the multiplicity increases for all values of $\eta$. 
The increase, however, is more pronounced for negative values of $\eta$, corresponding to particles going in the same direction as the Pb nucleus. 
Our goal is to understand what one can learn from this result, without resorting to a specific microscopic model of the collision dynamics (such as AMPT  \cite{Lin:2004en}, EPOS \cite{Pierog:2013ria}, or Angantyr \cite{Bierlich:2018xfw}). 

The concept of centrality originally refers to impact parameter $b$, not to $E_T$. 
More central collisions are those with smaller $b$, and the definition of the centrality fraction is $c=\pi b^2/\sigma$, where $\sigma$ is the inelastic cross section of the p+Pb collision.\footnote{We assume that $b$ is small enough that an inelastic collision occurs with probability $\sim 100\%$.}
A more central collision produces larger $E_T$ {\it on average\/}, but it is important to keep in mind that there are fluctuations around this average, which blur the relation between $E_T$ and centrality. 
In Pb+Pb collisions, fluctuations are small enough that the centrality estimated using $E_T$ has  $\sim 1\%$ accuracy. 
This justifies the use of $E_T$ as a centrality classifier~\cite{ALICE:2013hur,Das:2017ned}. 
This no longer holds in p+Pb collisions for two reasons. 
First, the system size is smaller, so that the fluctuations of $E_T$ around the average $\overline{E_T}(c)$ at fixed centrality are larger in relative value. 
Second, the average value $\overline{E_T}(c)$, depends less strongly on $c$ than in Pb+Pb collisions.\footnote{The variation of $\overline{E_T}(c)$ with $c$ can be reconstructed from the measured distribution of $E_T$ using a simple Bayesian analysis. The magnitude of the centrality dependence  is characterized by the dimensionless quantity $-d\ln \overline{E_T}/dc|_{c=0}$, whose value is $4$  in Pb+Pb collisions (Table I of Ref.~\cite{Yousefnia:2021cup}) and $1$ in p+Pb collisions (Sec.III C of Ref.~\cite{Pepin:2022jsd}). This implies that the centrality dependence is 4 times weaker in p+Pb collisions than in Pb+Pb collisions. Note that the cross section $\sigma$ of a Pb+Pb collision is larger by a factor $\sim 4$ than that of a p+Pb collision. Since $c\simeq\pi b^2/\sigma$, the values of $d\ln\overline{E_T}/db^2|_{b=0}$ are approximately {\it equal\/} for p+Pb and Pb+Pb collisions at the same collision energy.}
The fact that fluctuations hinder the determination of impact parameter in p+Pb collisions has always been recognized. 
Early analyses actually refrained from using the term ``centrality'' in this context~\cite{ALICE:2012eyl,CMS:2013jlh}, and used  ``event activity''~\cite{CMS:2013jsu} instead.  

There is a simple quantity which encapsulates the strength of the correlation between $E_T$ and impact parameter, which is the fraction of events above the ``knee'' of the distribution of $E_T$~\cite{Das:2017ned}. 
The knee is defined as the average value of $E_T$ for central collisions, that is, $\overline{E_T}(c=0)$, and its value can be accurately reconstructed. 
Events above the knee result from fluctuations of $E_T$ relative to this average value. 
They represent only $0.3\%$ of the total number of events in Pb+Pb collisions~\cite{Das:2017ned}, but as many as $7.5\%$ in p+Pb collisions~\cite{Rogly:2018ddx}. 
Events above the knee are referred to as {\it ultracentral\/} in the case of Pb+Pb collisions~\cite{CMS:2013bza,Gardim:2019brr,Samanta:2023amp,Nijs:2023bzv,CMS:2024sgx}.
We carry over this terminology to p+Pb collisions. 
The crucial point is that a large fraction of p+Pb collisions are ultracentral. 
Ultracentral p+Pb collisions essentially encompass {\it all\/} the high-multiplicity collisions. 

Our work is based on the hypothesis that the fluctuations of multiplicity ($dN_{ch}/d\eta$) and $E_T$ in ultracentral p+Pb collisions are dominated by quantum fluctuations, and that the variation of centrality (as defined by impact parameter) has a negligible effect. 
We assume that the joint distribution of $dN_{ch}/d\eta$ and $E_T$ in ultracentral collisions is essentially that at zero impact parameter, $b=0$. 
This implies that the variation of $dN_{ch}/d\eta$ (averaged over events) as a function of $E_T$ has nothing to do with centrality, but solely reflects the correlation between $dN_{ch}/d\eta$ and $E_T$, whose origin lies in the quantum processes leading to particle production. 
Since $dN_{ch}/d\eta$ and $E_T$ are measured in separate pseudorapidity windows, this correlation is the manifestation of a long-range rapidity correlation. 
 
\begin{figure}[ht]
\begin{center}
\includegraphics[width=\linewidth]{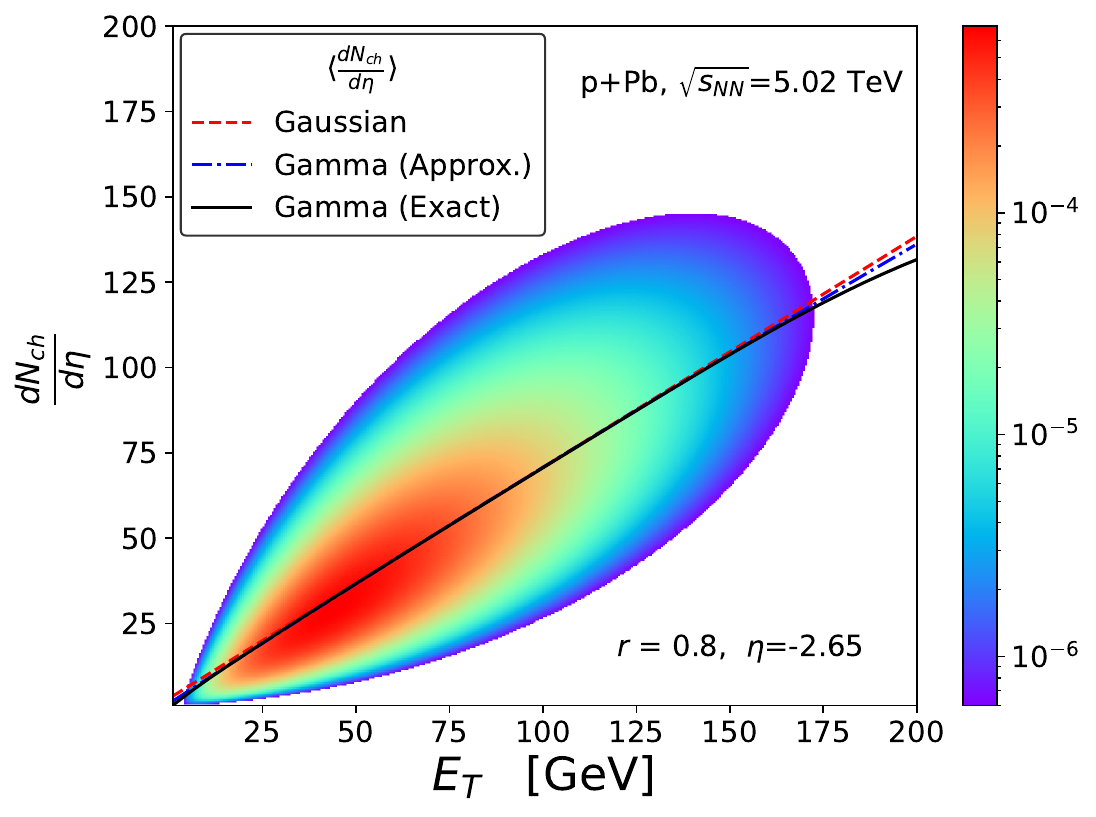} 
\end{center}
\caption{
Illustration of a correlated gamma distribution between transverse energy and multiplicity. 
For the multiplicity, we choose to represent $dN_{ch}/d\eta$ at $\eta=-2.65$, corresponding to the edge of the ATLAS acceptance in the Pb going side. 
The values of the parameters are extracted from ATLAS data as explained in the text. 
The only parameter which cannot be inferred from data is the Pearson correlation coefficient, for which we assume the value $r=0.8$. 
The lines display the average of $dN_{ch}/d\eta$ as a function of $E_T$, which is the quantity effectively measured by ATLAS. 
The full line is the exact result, the dashed line the Gaussian approximation (\ref{gaussianav2}), and the dash-dotted line the improved approximation 
 (\ref{gammaav2}), which is used throughout this paper. 
}
\label{fig:correlatedgamma}
\end{figure}

It was shown in Ref.~\cite{Pepin:2022jsd} that the joint distribution of $N_{ch}$ and $E_T$ at $b=0$ in the case of p+Pb collision is, to a good approximation, a correlated gamma distribution, whose definition is recalled in Appendix~\ref{s:gamma}, and which we denote by $P_\gamma(N_{ch},E_T)$. 
An example is depicted in Fig.~\ref{fig:correlatedgamma}. 
The correlated gamma distribution is a generic distribution, like the correlated Gaussian distribution, the main difference being that its support is the positive quadrant ($N_{ch}>0$, $E_T>0$), as opposed to the entire plane for the Gaussian distribution. 
The marginal distributions $P(N_{ch})$ and $P(E_T)$, obtained upon integration over one of the variables, are gamma distributions. 
The gamma distribution can be seen~\cite{Rogly:2018ddx} as a continuous version of the negative binomial distribution, which has been widely used to parametrize multiplicity fluctuations in high-energy proton-nucleus collisions~\cite{Gelis:2009wh,Tribedy:2011aa,Dumitru:2012yr,Bozek:2013uha}. 
Like the Gaussian distribution, the gamma distribution has only two parameters, the mean and the standard deviation. 
The correlated gamma distribution $P_\gamma(N_{ch},E_T)$ has five parameters, like the two-dimensional Gaussian distribution. 
These parameters are the mean values $\overline{N_{ch}}$ and $\overline{E_T}$, the standard deviations $\sigma_{N_{ch}}$ and $\sigma_{E_T}$, and the Pearson correlation coefficient $r$ between $N_{ch}$ and $E_T$: 
\begin{equation}
  \label{defpearson2}
r\equiv \frac{\langle N_{ch}E_T\rangle-\overline{N_{ch}} \ \overline{E_T}}{\sigma_{N_{ch}}\sigma_{E_T}}, 
\end{equation}

In this work, we assume that the probability distribution of $E_T$ and $dN_{ch}/d\eta$ at a given $\eta$ is also a correlated gamma distribution, for any value of $\eta$. 
To simplify expressions, we still use the notation  $N_{ch}$ to represent $dN_{ch}/d\eta$. 
The quantity measured by ATLAS in Fig.~\ref{fig:etadep} is the average value of $dN_{ch}/d\eta$ at fixed $E_T$, which we denote by $\langle N_{ch}|E_T\rangle$. 
It can be calculated as a function of the parameters of the gamma distribution, as will be shown below. 

\begin{figure}[h]
\begin{center}
\includegraphics[width=\linewidth]{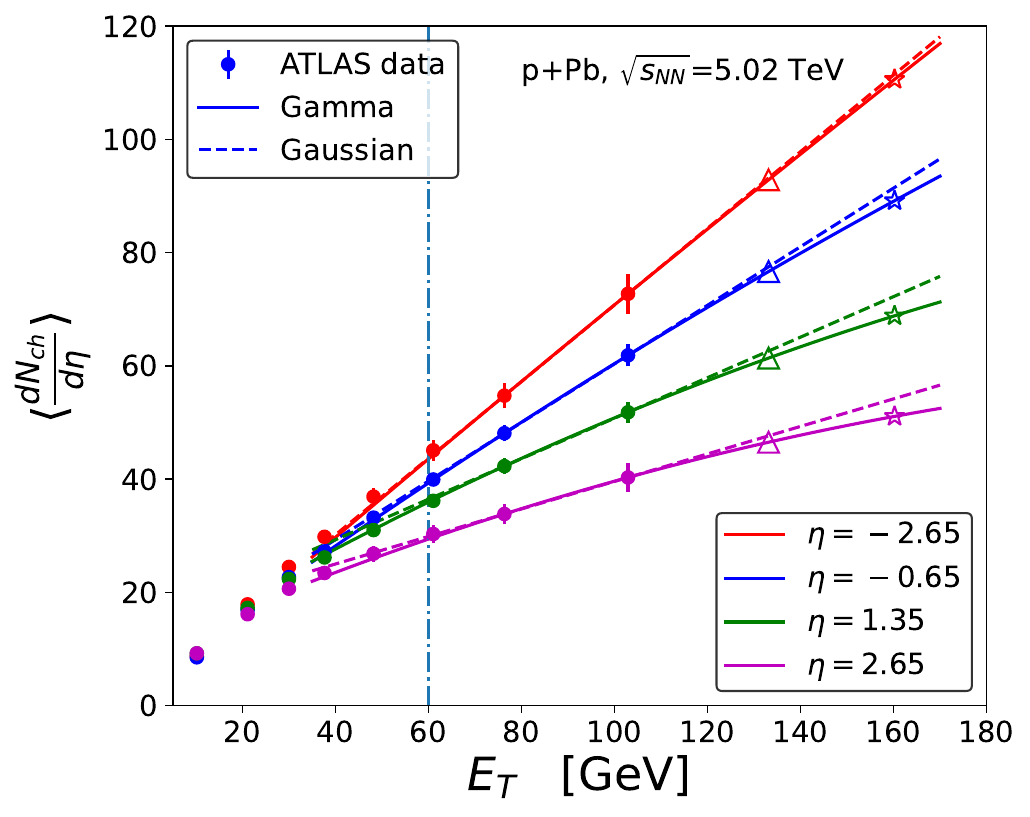} 
\end{center}
\caption{
Variation with $E_T$ of the average multiplicity (per unit pseudorapidity $\eta$) at fixed $E_T$, $\langle N_{ch}|E_T\rangle$, for four values of $\eta$. 
Full symbols correspond to ATLAS data which are the same as in Fig.~\ref{fig:etadep}, except for the change of variables from centrality percentile to $E_T$. 
The value of $E_T$ is taken at the center of the corresponding centrality window (e.g., for the $10-20\%$ window, we choose the value of $E_T$ corresponding to 15\% centrality). 
The vertical dash-dotted line indicates the position of the knee of the $E_T$ distribution, which we denote by $\overline{E_T}$. 
The dashed lines %shown for two extreme $\eta$ values,
are straight lines going through the two rightmost data point, correspond to the prediction of the Gaussian model  (\ref{gaussianav2}). 
Full lines correspond to the prediction from the correlated gamma distribution, Eq.~(\ref{gammaav2}), where the parameters are readjusted, so that the lines still go through the two rightmost data points. 
Triangles and stars are the resulting predictions for $0-0.1\%$ and $0-0.01\%$ centrality bins, which are  displayed in Fig.~\ref{fig:etadep} for all $\eta$. 
}
\label{fig:centralitydep}
\end{figure} 

We first consider the simpler case where  $N_{ch}$ and $E_T$ are distributed according to a correlated Gaussian distribution. 
The Gaussian distribution is simpler than the correlated gamma distribution, but both distributions coincide in the limit where $\sigma_{N_{ch}}\ll\overline{N_{ch}}$ and $\sigma_{E_T}\ll\overline{E_T}$. 
For the Gaussian distribution, a simple calculation gives (see Appendix~\ref{s:2dgaussian})
\begin{equation}
\label{gaussianav2}
\langle N_{ch}| E_T\rangle_{\text{Gaussian}}=\overline{N_{ch}}+\frac{r\sigma_{N_{ch}}}{\sigma_{E_T}}\left(E_T-\overline{E_T}\right). 
\end{equation}
This simple model thus predicts that the average multiplicity $\langle N_{ch}| E_T\rangle$ increases linearly with $E_T$, the increase being driven by  the Pearson correlation coefficient $r$. 
In order to compare this prediction with data, we plot in Fig.~\ref{fig:centralitydep} the same data as Fig.~\ref{fig:etadep}, but as a function of $E_T$ (as opposed to the ``centrality percentile''), for selected values of $\eta$. 
One sees that for all values of $\eta$, the increase of $dN_{ch}/d\eta$ is approximately linear in $E_T$ for the largest values of $E_T$, as predicted by Eq.~(\ref{gaussianav2}). 
%However, there is a noticeable {\it non-linear departure} for the bins with large positive $\eta$, as could be seen by eye from Fig.~\ref{fig:centralitydep}. We will come back to this shortly.

Let us discuss how the parameters entering Eq.~(\ref{gaussianav2}) can be obtained from data. 
The mean and standard deviation of $E_T$ at $b=0$, $\overline{E_T}$ and $\sigma_{E_T}$, can be reconstructed~\cite{Rogly:2018ddx} through a Bayesian analysis of the distribution of $E_T$, $dN/dE_T$, which is measured by ATLAS. 
The idea is to assume that the distribution of $E_T$ at fixed $b$ is a gamma distribution, so that one fits $dN/dE_T$ as a superposition of gamma distributions. 
This analysis has been carried out in Sec.III C of Ref.~\cite{Pepin:2022jsd}, and the following values have been obtained, which we use throughout this paper: $\overline{E_T}= 60\pm 2$~GeV and $\sigma_{E_T}= 24\pm 2$~GeV . 

The other parameters are $\overline{N_{ch}}$, $\sigma_{N_{ch}}$, and $r$, which are functions of  $\eta$.  
According to Eq.~(\ref{gaussianav2}), $\overline{N_{ch}}$ is the value of $\langle N_{ch}| E_T\rangle_{\text{Gaussian}}$ at  $E_T=\overline{E_T}$, while the slope of the lines in Fig.~\ref{fig:centralitydep}  gives access to $r\sigma_{N_{ch}}$. 
Note that one cannot reconstruct $r$ and $\sigma_{N_{ch}}$ separately, but only the product. 
This can be understood as follows:
One measures a correlation between $N_{ch}$ and $E_T$, and $r\sigma_{N_{ch}}$ is proportional to this correlation, according to Eq.~(\ref{defpearson2}). 

%The non-linearity of $E_T$ dependence of $dN_{ch}/d\eta$, seen in Fig.~\ref{fig:centralitydep}, definitely demands an improvement over the Gaussian approximation. 
We now consider the more realistic case where  $N_{ch}$ and $E_T$ are distributed according to a correlated  gamma distribution. 
The leading correction to Eq.~(\ref{gaussianav2}) is (see derivation in Appendix~\ref{s:2dgamma}):
\begin{equation}
\begin{aligned}
\langle &N_{ch}| E_T\rangle_{\text{Gamma}}=\overline{N_{ch}}+\frac{r\sigma_{N_{ch}}}{\sigma_{E_T}}\left(E_T-\overline{E_T}\right)+\cr
&\frac{r\sigma_{N_{ch}}}{3}\left(\frac{\sigma_{E_T}}{\overline{E_T}}-\frac{r\sigma_{N_{ch}}}{\overline{N_{ch}}}\right)\left[1-\frac{\left(E_T-\overline{E_T}\right)^2}{\sigma_{E_T}^2}\right]. 
\end{aligned}
\label{gammaav2}
\end{equation}
This can be decomposed as 
\begin{equation}
\begin{aligned}
\langle N_{ch}| E_T\rangle_{\text{Gamma}}=&\langle N_{ch}| E_T\rangle_{\text{Gaussian}}+ \cr
&C\left[1-\frac{\left(E_T-\overline{E_T}\right)^2}{\sigma_{E_T}^2}\right], 
\end{aligned}
\label{gammaav3}
\end{equation}
so that an additional term appears, which is quadratic in $E_T$, with a  coefficient
\begin{equation}
\begin{aligned}
\ C%(\overline{N_{ch}},r\sigma_{N_{ch}}) 
\equiv \frac{r\sigma_{N_{ch}}}{3}\left(\frac{\sigma_{E_T}}{\overline{E_T}}-\frac{r\sigma_{N_{ch}}}{\overline{N_{ch}}}\right).
\end{aligned}
\label{coeffquadr}
\end{equation}
%gives a positive number for all values of $\eta$. 
%The variation with $E_T$ is no longer linear, but involves a quadratic term. 
Similar to Eq.~(\ref{gaussianav2}), $r$ and $\sigma_{N_{ch}}$ only appear through the combination $r\sigma_{N_{ch}}$. 
Equation~(\ref{gammaav2}) is more precise than Eq.~(\ref{gaussianav2}), and this improvement comes at no additional cost, since the number of parameters is the same in both equations. 
For a given pseudorapidity window, there are two unknown quantities, $\overline{N_{ch}}$ and $r\sigma_{N_{ch}}$. 
They can be obtained by measuring  $\langle N_{ch}| E_T\rangle$ for two different values of $E_T$, and solving Eq.~(\ref{gammaav2}). 
Since our model is tailored for high-multiplicity collisions, which are collisions with the highest $E_T$, we determine $\overline{N_{ch}}$ and $r\sigma_{N_{ch}}$ using the two rightmost data points in Fig.~\ref{fig:centralitydep}. 

The first question is whether Eq.~(\ref{gammaav2}) differs significantly from Eq.~(\ref{gaussianav2}). 
The lines in Fig.~\ref{fig:correlatedgamma} compare these two equations for a specific value of $\eta$,  corresponding to the edge of the ATLAS acceptance in the direction of the Pb nucleus. 
Even though the distribution represented in color (Fig.~\ref{fig:correlatedgamma}) is strongly asymmetric, so that one would expect the Gaussian approximation to be poor, it works quite well for this particular  observable. 
Note that the result obtained using Eq.~(\ref{gammaav2}), where only the leading correction to the Gaussian is taken into account, is very close to the exact result (shown as a solid line) obtained by integrating numerically the correlated gamma distribution. 

The curves corresponding to Eq.~(\ref{gammaav2}) lie below the Gaussian approximation in Eq.~(\ref{gaussianav2}) for large $E_T$, in Fig.~\ref{fig:centralitydep}. 
This means that the additional quadratic term in Eq.~(\ref{gammaav3}) is negative or, equivalently, that $C$ is positive. 
Using Eq.~(\ref{coeffquadr}), this implies:   
\begin{equation}
\frac{r\sigma_{N_{ch}}}{\overline{N_{ch}}} < \frac{\sigma_{E_T}}{\overline{E_T}}. 
\end{equation} 
This inequality holds for all values of $\eta$, for three reasons which we briefly outline. 
First, dynamical fluctuations of $E_T$ have been shown to be larger than those of $N_{ch}$ in relative value~\cite{Pepin:2022jsd}, that is, $\sigma^{\rm dyn}_{N_{ch}}/\overline{N_{ch}}<\sigma^{\rm dyn}_{E_T}/\overline{E_T}$, where the superscript ``dyn'' refers to the dynamical contribution to the fluctuations~\cite{Pruneau:2002yf}. 
Second, statistical fluctuations increase $\sigma_{E_T}$ relative to the dynamical component $\sigma^{\rm dyn}_{E_T}$. This is however a small increase, of order $3\%$.\footnote{The statistical contribution to the variance $\sigma_{E_T}^2$ is estimated to be $\sim 6\%$ in Sec. IV A of Ref.~\cite{Pepin:2022jsd}.}
Third, the Pearson correlation coefficient $r$ between $N_{ch}$ and $E_T$ is smaller than unity as a result of the rapidity decorrelation~\cite{CMS:2015xmx}. 
One expects that $r$ decreases as the rapidity gap increases (see, e.g., Fig.~8 of Ref.~\cite{Pepin:2022jsd}). 
Since $E_T$ is measured in the pseudorapidity window $-4.9<\eta<-3.1$,  $r$ should decrease as $\eta$ increases. As a consequence, the value of the coefficient $C$ increases for larger positive $\eta$ values, making the negative suppression more pronounced as seen from Fig.~\ref{fig:centralitydep}. 

Once the parameters $\overline{N_{ch}}$ and $r\sigma_{N_{ch}}$ are determined from two values of $E_T$, Eq.~(\ref{gammaav2}) can be used to obtain the average multiplicity for any other value of $E_T$. 
This corresponds to the solid lines in Fig.~\ref{fig:centralitydep}.  
One sees that the nonlinear corrections become sizable for large values of $E_T$. 
It would be interesting to see if experimental data confirm this trend. 

The full comparison between our model and ATLAS data is displayed in Fig.~\ref{fig:etadep}. 
The parameters are adjusted so as to match the data in the centrality bins $0-1\%$ and $1-5\%$, for all values of $\eta$.
Therefore, the model is on top of data for these two bins by construction. 
The next bin, $5-10\%$, is well reproduced by our model. 
Differences between model and data increase as the centrality percentile increases. 
This can be easily understood since we have neglected the effect of the variation of impact parameter. 
More specifically, we have assumed $b\approx 0$. 
As argued above, this is a reasonable approximation above the knee of the distribution, that is, below 7.5\% centrality. 
But it becomes gradually worse for less central collisions. 
Agreement is however reasonable all the way up to 20\% centrality. 
We also display our prediction for $0-0.1\%$ and $0-0.01\%$ centralities, corresponding to $E_T\simeq133$~GeV and $E_T\simeq160$~GeV, respectively. 

\begin{figure}[h]
\begin{center}
\includegraphics[width=\linewidth]{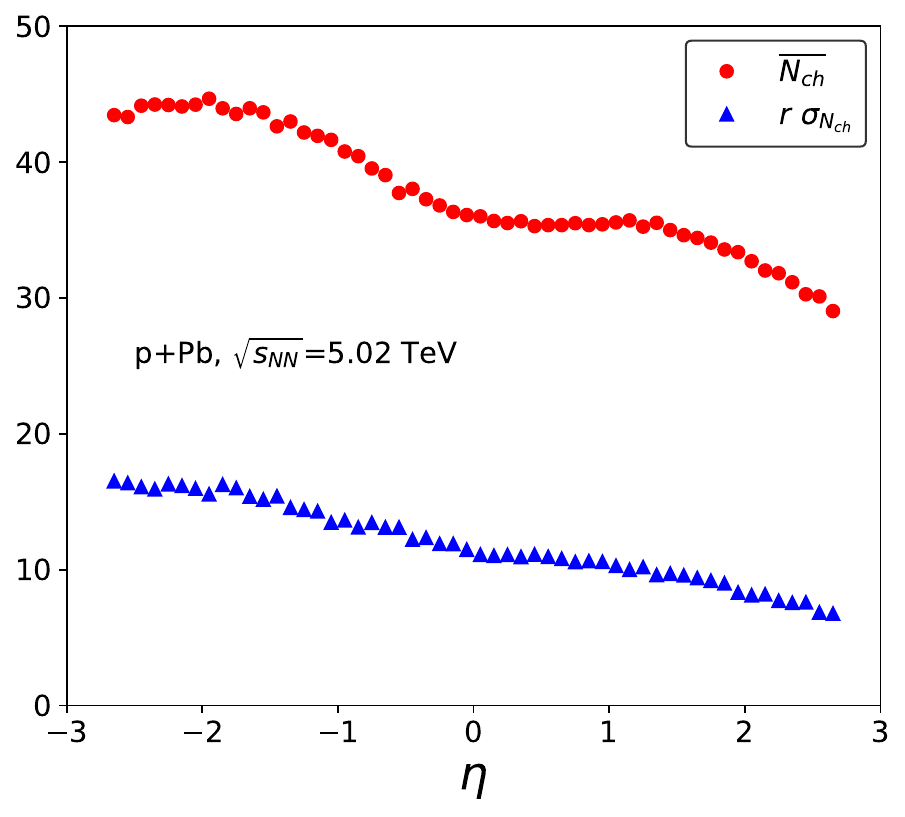} 
\end{center}
\caption{
Pseudorapidity dependence of the average multiplicity in $b=0$ collisions, $\overline{N_{ch}}$, and of $r\sigma_{N_{ch}}$, representing its correlation with $E_T$ (see Eq.~(\ref{defpearson2})). 
These values are obtained by fitting ATLAS data for $0-1\%$ and $1-5\%$ centrality using Eq.~(\ref{gammaav2}). 
}
\label{fig:fitparam}
\end{figure}

We have shown that for events with large $E_T$, the ``centrality dependence'' of the multiplicity is not driven by the variation of centrality, but by the correlation between multiplicity and $E_T$, that is, by a long-range rapidity correlation. 
We now discuss which information one obtains about this correlation, and how future analyses may improve our knowledge of rapidity correlations. 
The output of our analysis is the value of the two parameters entering Eq.~(\ref{gammaav2}), namely, $\overline{N_{ch}}$ and $r\sigma_{N_{ch}}$. 
They are displayed in Fig.~\ref{fig:fitparam}.   
Both decrease as a function of rapidity, the dip around $\eta=0$ being due to the transformation between pseudorapidity and rapidity~\cite{PHENIX:2015tbb}. 
One expects the correlation coefficient $r$ to decrease as a function of $\eta$, since the rapidity gap with the calorimeter increases. 
However, one cannot disentangle $r$ from $\sigma_{N_{ch}}$ with these data. 

In order to disentangle the correlation $r$ from the fluctuation $\sigma_{N_{ch}}$, one can repeat the analysis with a different centrality classifier. 
Instead of binning events in $E_T$, ATLAS could bin them according to the multiplicity in a subset of the central detector (which covers $-2.65<\eta<2.65$), or according to the transverse energy in the forward calorimeter $3.1<\eta<4.9$. 
Selecting the 5\% ``most central'' events with this new centrality classifier, and repeating the above analysis, one will obtain another set of values of $\overline{N_{ch}}$ and $r\sigma_{N_{ch}}$ as functions of $\eta$. 
Both $\overline{N_{ch}}$ and $\sigma_{N_{ch}}$ should reflect intrinsic properties of the distribution of $N_{ch}$ at $b=0$, and should not depend on the centrality classifier. 
If the hypothesis made in this paper are correct, then the values of $\overline{N_{ch}}$ displayed in Fig.~\ref{fig:fitparam} should be independent of the centrality classifier.\footnote{One should however exclude the values of $\eta$ which are used to determine the centrality so as to avoid self-correlations.}   
This can serve as a validation of the method. 
If this is verified, then any difference in the value of $r\sigma_{N_{ch}}$ can safely be attributed to a different value of $r$, so that the analysis would give direct access to the variation of $r$ with the rapidity gap. 
This phenomenon is closely correlated to the  longitudinal decorrelation of azimuthal correlations, which have been much studied in the last decade~\cite{CMS:2015xmx,ATLAS:2017rij,Bozek:2017qir,Pang:2018zzo,ATLAS:2020sgl}. 

Slightly different centrality classifiers have already been tried by ATLAS, in order to assess the robustness of their results. 
In particular, they have repeated the analysis by using only half of the backward calorimeter (covering the window $-4.9<\eta<-4.0$), as opposed to the whole calorimeter ($-4.9<\eta<-3.1$) used for the nominal analysis. 
They find that for the 1\% most central events, the average value of $dN_{ch}/d\eta$ is reduced to $97\%-98\%$ of the nominal value  (Fig. 9 of \cite{ATLAS:2015hkr}). 
Since the $\eta$ window is similar in both cases, one does not expect any significant modification from rapidity correlations. 
It is however instructive to investigate whether the reduction observed by ATLAS is compatible with our model. 
For this centrality bin, representing the rightmost data  point in Fig.~\ref{fig:centralitydep}, $E_T\simeq 103$~GeV. 
Equation (\ref{gaussianav2}) (which is accurate enough for this discussion) then gives
\begin{equation}
\label{rightmostbin}
    \langle N_{ch}| E_T\rangle\simeq \overline{N_{ch}}+1.8 \ r\sigma_{N_{ch}}. 
\end{equation}
As discussed above, the first term in the right-hand side is independent of the centrality classifier, hence the dependence should solely come from the second term, more precisely from $r$. 
Using the values in Fig.~\ref{fig:fitparam}, one sees that the second term contributes between $\sim 41\%$ down to $30\%$ of the total (the fraction decreasing as $\eta$ increases). 
The decrease of $dN_{ch}/d\eta$ observed by ATLAS then implies that the second term is smaller by $\sim 7\%$ than for the nominal measurement. 
This reduction can be attributed to statistical fluctuations. 
By using half of the calorimeter, the transverse energy is roughly divided by 2. 
This holds both for the average value and the width of dynamical fluctuations~\cite{Pruneau:2002yf}. 
On the other hand, the statistical contribution to $\sigma_{E_T}$, coming from the random hadronization process, is only divided by $\sqrt{2}$.  
These statistical fluctuations do not induce any correlation with $N_{ch}$. 
Hence, larger statistical fluctuations imply smaller $r$. 
The decrease by $7\%$ observed by ATLAS with half of the calorimeter implies that statistical fluctuations are increased by 7\% relative to the nominal analysis. 
This is compatible with the estimate provided in Sec. IV A of Ref.~\cite{Pepin:2022jsd}, that statistical fluctuations contribute by $6\%$ to the variance of $E_T$. 
This constitutes a first non-trivial validation of our approach.  

In conclusion, we have shown that the so-called centrality dependence of the multiplicity actually reflects its correlation with the centrality classifier. 
For the $\sim 10\%$ most central collisions, one can even ignore the variation of impact parameter. 
The fluctuations of the multiplicity and of the centrality classifier are both dominated by quantum fluctuations. 
Their probability distribution is well approximated by a correlated gamma distribution, from which the multiplicity and the centrality classifier are sampled. 
We have argued that by repeating this analysis with another centrality classifier, covering a different rapidity range, one will be able to extract direct information about the rapidity decorrelation in the process of particle production. 

%multiplicity decorrelation and flow decoherence?

\begin{acknowledgments}
R. S. is supported by the Polish grant NCN grant PRELUDIUM BIS: 2019/35/O/ST2/00357. R. S. also acknowledges the STRONG-HFHF-2023 workshop organised in Giardini Naxos, Sicily, Italy for helpful discussions during the workshop. 
\end{acknowledgments}

\appendix
\section{Correlated Gaussian and gamma distributions}
\label{s:gamma}

In this Appendix, we define the correlated gamma distribution of two variables $(N_{ch},E_T)$, which we use to model their fluctuations. 
The goal is to derive the expression (\ref{gammaav2}) of the average value of $N_{ch}$ at fixed $E_T$, on which the calculations of this article are based. 
We start with the gamma distribution of a single variable $x$. 
We derive, in Sec.~\ref{s:onevariable}, the leading correction to the central limit theorem, on which Eq.~(\ref{gammaav2}) is based. 
In Sec.~\ref{s:mapping}, we show that a Gaussian distribution can be transformed into a gamma distribution through a simple change of variables.  
We then move on to the two-variable case  $(x_1,x_2)$. 
We recall standard results on the correlated Gaussian distribution, in Sec.~\ref{s:2dgaussian}, and then generalize them to a correlated gamma distribution in Sec.~\ref{s:2dgamma}. 

\subsection{gamma versus Gauss: leading corrections}
\label{s:onevariable}

The gamma distribution is 
\begin{equation}
\label{gamma}
P_\gamma(x)=\frac{1}{\Gamma(k)\theta^k}x^{k-1} e^{-x/\theta}.
\end{equation}
The mean and the standard deviation are given by:
\begin{eqnarray}
  \label{meanvariance}
\bar  x &=&\theta  k\cr
\sigma &=&\theta\sqrt{k}.  
\end{eqnarray}
Fluctuations are small if $\sigma\ll \bar x$ or, equivalently, if $k\gg 1$.
The gamma distribution reduces to a Gaussian in this limit, according to the central limit theorem. 
Here we derive the leading correction to this limit.  
%I then discuss the case of the correlated gamma distribution.

We now introduce the Gaussian distribution $P_G(x)$ with the same mean and variance as $P_\gamma(x)$:
\begin{eqnarray}
\label{gaussian}
P_G(x)&=&\frac{1}{\sigma\sqrt{2\pi}} \exp\left(-\frac{(x-\bar x)^2}{2\sigma^2}\right)\cr
&=&\frac{1}{\theta\sqrt{2\pi k}} \exp\left(-\frac{(x-k\theta)^2}{2k\theta^2}\right).
\end{eqnarray}
The comparison between these two distributions must be done within a few $\sigma$ around the mean $\bar x$, since the probability vanishes outside this region. 
We therefore decompose $x$ as 
\begin{equation}
\label{decomposition}
x=\bar x+\epsilon=k\theta+\epsilon,
\end{equation}
where $\epsilon$ is of order $\sigma=\theta\sqrt{k}$. 
The relative difference between $P_\gamma(x)$ and $P_G(x)$, when small, is the difference between the logarithms. 
Using Eqs.~(\ref{gamma}), (\ref{gaussian}) and (\ref{decomposition}), one obtains: 
\begin{eqnarray}
\label{lngamma}
\ln P_\gamma(\bar x+\epsilon)&=&-\frac{\epsilon}{\theta}+(k-1)\ln\left(1+\frac{\epsilon}{k\theta}\right)\cr
\ln P_G(\bar x+\epsilon)&=&-\frac{\epsilon^2}{2k\theta^2},
\end{eqnarray}
up to additive constants independent of $\epsilon$. 
For $k\gg 1$, $\epsilon/k\theta$ is of order $k^{-1/2}$, and one can expand the logarithm in powers of $\epsilon/k\theta$:
\begin{equation}
\label{lngamma2}
\ln P_\gamma(\bar x+\epsilon)=-\frac{\epsilon^2}{2k\theta^2}-
\frac{\epsilon}{k\theta}+\frac{\epsilon^3}{3k^2\theta^3}+{\cal O}(k^{-1}), 
\end{equation}
where the first term is of order unity and the next two terms are the leading corrections, of order $k^{-1/2}$. 
Taking the difference between Eqs.~(\ref{lngamma2}) and the second line of  (\ref{lngamma}), exponentiating and expanding to leading order in the correction, one obtains: 
\begin{eqnarray}
\label{gammaversusgauss}
P_\gamma(\bar x+\epsilon)&=&P_G(\bar x+\epsilon)\left(1-
\frac{\epsilon}{k\theta}+\frac{\epsilon^3}{3k^2\theta^3}\right) \cr
&=&P_G(\bar x+\epsilon)\left(1-
\frac{\epsilon}{\bar x}+\frac{\epsilon^3}{3\bar x\sigma^2}\right)
\end{eqnarray}
The correction terms are both odd in $\epsilon$, therefore, they integrate to 0, which means that there is no need for a correction to account for the normalization, i.e., for the additive constants that were ignored in the logarithm. 

Note that Eq.~(\ref{gammaversusgauss}) is a leading-order correction which does not verify the same mathematical  properties as the gamma distribution:
$P_\gamma(\bar x+\epsilon)$ becomes negative for large negative $\epsilon$, but the corresponding values of the probability are exponentially suppressed. 
Its support is the real axis, as opposed to the positive semi-axis for the gamma distribution, but the (negative) values of the probability are also exponentially suppressed for $\bar x+\epsilon<0$. 

\begin{figure}[h]
\begin{center}
\includegraphics[width=\linewidth]{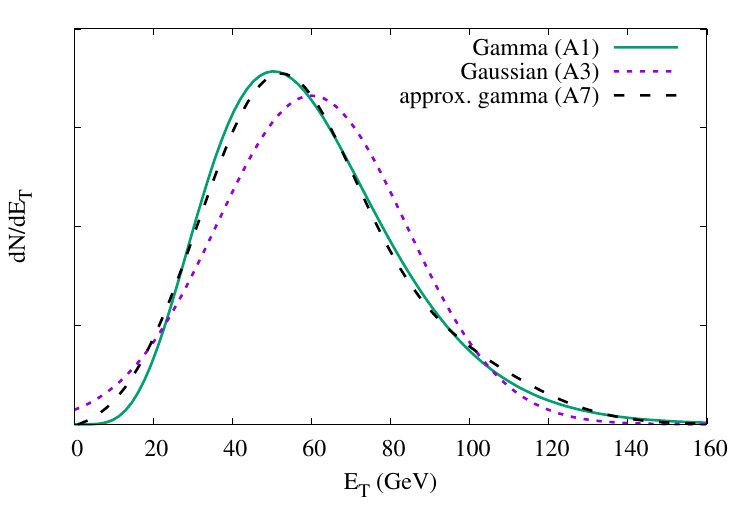} 
\end{center}
\caption{
Comparison between Eqs.~(\ref{gamma}),  (\ref{gaussian}) and  (\ref{gammaversusgauss}). 
The mean and the standard deviation are $60$~GeV and $24$~GeV, corresponding to the distribution of $E_T$ in p+Pb collisions at $b=0$~\cite{Pepin:2022jsd}. 
}
\label{fig:gamma}
\end{figure} 
Fig.~\ref{fig:gamma} displays a comparison between the Gaussian distribution, the gamma distribution, and the approximation (\ref{gammaversusgauss}) to the gamma distribution. 
The mean and standard deviation are identical to those of the distribution of $E_T$ in Fig.~\ref{fig:correlatedgamma}. 
One sees that Eq.~(\ref{gammaversusgauss}) represents a very reasonable approximation for typical parameter values used in this work. 

\subsection{Mapping the Gaussian onto a gamma distribution}
\label{s:mapping}

We now derive the change of variables $x\to x'$ such that if the distribution of $x$ is $P_G(x)$, then the distribution of $x'$ is $P_\gamma(x')$. 
This is achieved by matching the cumulative distributions: 
\begin{equation}
\label{cumulative}
\int_{-\infty}^{x}P_G(t)dt=\int_{0}^{x'} P_\gamma(t)dt.
\end{equation} 
We derive this mapping in the limit where the relative difference between $P_\gamma(x)$ and $P_G(x)$ is small, and Eq.~(\ref{gammaversusgauss}) applies. 
This in turn implies that $x'-x$ is small, so that the right-hand side of Eq.~(\ref{cumulative}) can be decomposed into: 
\begin{eqnarray}
\label{cumulativesmall}
\int_{-\infty}^{x'} P_\gamma(t)dt&=&\int_{-\infty}^x P_\gamma(t)dt+(x'-x)P_\gamma(x)\cr &\simeq& \int_{-\infty}^x P_\gamma(t)dt+(x'-x)P_G(x),
\end{eqnarray}
where, in the last line, we have used the fact that $x'-x$ is a small correction so that the factor in front can be evaluated to order 0 in the correction. 
Note also that the lower integration limit in the right-hand side has been changed from $0$ to $-\infty$, for the reasons explained at the end of Sec.~\ref{s:onevariable}. 
Eqs.~(\ref{cumulative}) and (\ref{cumulativesmall}) give:
\begin{equation}
\label{mapping0}
x'-x=\frac{1}{P_G(x)}\int_{-\infty}^{x} \left(P_G(t)-P_\gamma(t)\right) dt.
\end{equation}
Changing variables from $x$ to $\epsilon=x-\bar x$ and using Eq.~(\ref{gammaversusgauss}), we then obtain
\begin{eqnarray}
\label{mapping1}
x'-x&=&\frac{1}{\exp\left(-\epsilon^2/2\sigma^2\right)}\int_{-\infty}^{\epsilon} \left(\frac{t}{\bar x}-\frac{t^3}{3\bar x\sigma^2}\right)
\exp\left(-\frac{t^2}{2\sigma^2}\right)\cr
&=&\frac{\epsilon^2-\sigma^2}{3\bar x}.
\end{eqnarray}
Eventually, the mapping of the Gaussian distribution onto the gamma distribution is implemented, to leading order in $k^{-1/2}$, by the simple change of variables
\begin{equation}
\label{mapping}
x'=x+\frac{\epsilon^2-\sigma^2}{3\bar x}, 
\end{equation}
where $\epsilon=x-\bar x$. 
%Note that the integral of $(x'-x)P_G(x)$ is 0, which is a consistency check since the means of the two distributions have been assumed to be equal. 
Eq.~(\ref{mapping}) shows that $x'>x$ for $\epsilon<-\sigma$ and for $\epsilon>\sigma$. 
This means that the tails of the distribution are pushed to the right on both sides (Fig.~\ref{fig:gamma}), which generates a right (positive) skew.

\subsection{Correlated Gaussian distribution}
\label{s:2dgaussian}

We consider two variables $x_1$ and $x_2$, whose distribution $P_G(x_1,x_2)$ is a correlated Gaussian.
We denote by $\bar x_i$ and $\sigma_i$ the mean and standard deviation of $x_i$,  and by $\epsilon_i=x_i-\bar x_i$ the deviation from the mean. 
The correlated Gaussian  is defined by: 
\begin{equation}
\label{corrgaussian}
\ln P_G(x_1,x_2)=-\frac{1}{1-r^2}\left(\frac{\epsilon_1^2}{2\sigma_1^2}+\frac{\epsilon_2^2}{2\sigma_2^2}-\frac{r\epsilon_1\epsilon_2}{\sigma_1\sigma_2}\right),
\end{equation}
up to an additive constant which ensures that $P_G(x_1,x_2)$ is normalized to unity. 
In this expression, $r$ denotes the Pearson correlation coefficient:
\begin{equation}
  \label{defpearson}
r\equiv \frac{\langle \epsilon_1\epsilon_2\rangle}{\sigma_1\sigma_2}, 
\end{equation}
where angular brackets in the numerator denote the expectation value, namely, the integral over $x_1$ and $x_2$ weighted with $P_G(x_1,x_2)$. 
Note that the marginal distributions $P_G(x_1)\equiv\int P_G(x_1,x_2)dx_2$ (and similarly for $x_2$) are Gaussian distributions of the type (\ref{gaussian}). 

For fixed $\epsilon_2$, inspection of Eq.~(\ref{corrgaussian}) shows that the probability distribution of $\epsilon_1$ is a Gaussian distribution. 
The average value of $\epsilon_1$ for  fixed $\epsilon_2$, which we denote by $\langle \epsilon_1| \epsilon_2\rangle$,  is the center of this Gaussian, which is the maximum of the right-hand side. 
Differentiating Eq.~\ref{corrgaussian} with respect to $\epsilon_1$ at fixed $\epsilon_2$, one immediately obtains
\begin{equation}
\label{gaussianav}
\langle \epsilon_1| \epsilon_2\rangle=\frac{r\sigma_1}{\sigma_2} \epsilon_2.
\end{equation}
This is equivalent to Eq.~(\ref{gaussianav2}), up to the change of notations $(x_1,x_2)\to (N_{ch},E_T)$. 

\subsection{Correlated gamma distribution}
\label{s:2dgamma}

We now define a correlated gamma distribution of two variables $(x'_1,x'_2)$ from the correlated Gaussian distribution, by mapping  $x_i$ onto $x'_i$ according to Eq.~(\ref{mapping})~\cite{Pepin:2022jsd}. 
Then the distributions of $x'_1$ and $x'_2$ are gamma distributions, and the correlation between $x_1$ and $x_2$ induces a correlation between $x'_1$ and $x'_2$.
We define the correlated gamma distribution $P_\gamma(x'_1,x'_2)$ as the joint distribution of $x'_1$ and $x'_2$.\footnote{This trick was suggested to JYO by Kristjan Gulbrandsen from the Niels Bohr Institute in Copenhagen.}

We first show that to leading order, the Pearson correlation coefficient (\ref{defpearson}) is identical for the two distributions. 
Decomposing $x'_i=\bar x_i+\epsilon'_i$, Eq.~(\ref{mapping}) gives:
\begin{equation}
\label{mapping2}
\epsilon'_i=\epsilon_i+\frac{\epsilon_i^2-\sigma_i^2}{3\bar x_i}, 
\end{equation}
where the last term on the right-hand side is the leading-order correction. 
Then one sees that $\langle  \epsilon'_1 \epsilon'_2\rangle=\langle  \epsilon_1 \epsilon_2\rangle$, since the cross-terms 
$\langle\epsilon_1(\epsilon_2^2-\sigma_2^2)\rangle$ and $\langle\epsilon_2(\epsilon_1^2-\sigma_1^2)\rangle$ only involve odd moments, which vanish for a Gaussian distribution. 

The explicit expression of $P_\gamma(x'_1,x'_2)$ is obtained by matching the probability:
\begin{equation}
\label{defpgamma}
P_\gamma(x'_1,x'_2)dx'_1dx'_2= P_G(x_1,x_2)dx_1 dx_2 . 
\end{equation}
Replacing $x'_i$ with $x_i+\delta_i$, and using $dx'_i/dx_i=1+d\delta_i/dx_i$, one obtains to first order in $\delta_i$
\begin{eqnarray}
  \label{defpgamma2}
  \frac{P_\gamma(x_1,x_2)}{P_G(x_1,x_2)}-1&=&-\frac{\partial\ln P_G(x_1,x_2)}{\partial x_1}\delta_1-\frac{d\delta_1}{dx_1}\cr&&\cr &&+(1\leftrightarrow 2), 
\end{eqnarray}
where $(1\leftrightarrow 2)$ means that the right-hand side must be symmetrized. 
Inserting Eqs.~(\ref{corrgaussian}) and (\ref{mapping}) into Eq.~(\ref{defpgamma2}), one obtains the following explicit expression: 
\begin{eqnarray}
\label{defpgamma3}
 \frac{P_\gamma(x_1,x_2)}{P_G(x_1,x_2)}-1&=&\left(\frac{\epsilon_1}{\sigma_1}-\frac{r\epsilon_2}{\sigma_2}\right)\frac{\epsilon_1^2-\sigma_1^2}{3(1-r^2)\sigma_1\bar x_1} -\frac{2\epsilon_1}{3\bar x_1}\cr &&+(1\leftrightarrow 2). 
\end{eqnarray}
Using this expression of $P_\gamma(x_1,x_2)$, multiplying with $\epsilon_1$, and integrating over $\epsilon_1$, one obtains the following expression of the average value of $\epsilon_1$ for fixed $\epsilon_2$:  
\begin{equation}
\label{gammaav}
\langle \epsilon_1| \epsilon_2\rangle=\frac{r\sigma_1}{\sigma_2} \epsilon_2+\frac{r\sigma_1}{3}\left(\frac{\sigma_2}{\bar x_2}-\frac{r\sigma_1}{\bar x_1} \right)\left(1-\frac{\epsilon_2^2}{\sigma_2^2}\right). 
\end{equation}
The first term on the right-hand side is the result for the Gaussian distribution, Eq.~(\ref{gaussianav}), and the second term is the leading-order correction for the gamma distribution. 
Note that after averaging over $\epsilon_2$, both terms vanish since $\langle\epsilon_2\rangle=0$ and $\langle\epsilon_2^2\rangle=\sigma_2^2$, and one recovers $\langle\epsilon_1\rangle=0$, as one should. 

Equation~(\ref{gammaav}) is equivalent to Eq.~(\ref{gammaav2}), up to the change of notations $(x_1,x_2)\to (N_{ch},E_T)$.

%In the limiting case $r\to 1$, the support of $(N'_1,N'_2)$ is a straight line.
%The support of the transformed variables $(N_1,N_2)$ is still a line, but it is not a straight line unless $k_1=k_2$.
%The line always goes through the origin $(0,0)$. 

\end{document}